\documentclass[aps,prl,reprint,superscriptaddress]{revtex4-2}

\usepackage{amsmath} 
\usepackage{graphicx} 
\usepackage{float}
\usepackage{helvet}
\usepackage{makecell}
\usepackage{multirow}
\usepackage{siunitx}
\usepackage{booktabs} 
\usepackage{makecell} 

\renewcommand{\arraystretch}{1.25}

\usepackage{soul}
\usepackage[dvipsnames]{xcolor}

\usepackage[version=4]{mhchem}

\usepackage{comment}

\begin{document}

\title{A Defect-Free Model of Amorphous Silicon with Pristine Electronic Structure}

\author{Louise A. M. Rosset}
\affiliation{Inorganic Chemistry Laboratory, Department of Chemistry, University of Oxford, Oxford, UK}

\author{Chinonso Ugwumadu}
\affiliation{Quantum \& Condensed Matter (T-4) Group, Los Alamos National Laboratory, Los Alamos, NM, USA}

\author{Stephen R. Elliott}
\affiliation{Physical and Theoretical Chemistry Laboratory, Department of Chemistry, University of Oxford, Oxford, UK}

\author{David A. Drabold}
\affiliation{Department of Physics and Astronomy, Ohio University, Athens, OH, USA}

\author{Volker~L.~Deringer}
\email{volker.deringer@chem.ox.ac.uk}
\affiliation{Inorganic Chemistry Laboratory, Department of Chemistry, University of Oxford, Oxford, UK}

\begin{abstract}
Amorphous silicon ($a$-Si) is understood to be the canonical continuous random network material, ideally defined by fully fourfold coordination. Here, we show that a defect-free (`ideal') model of $a$-Si from machine-learning-driven molecular-dynamics simulations [L.\ A.\ M. Rosset et al., Nat.\ Commun.\ {\bf 16}, 2360 (2025)], subsequently evaluated with hybrid-level density-functional theory computations, can accurately reproduce the experimentally observed electronic bandgap. We compare this model with one resulting from the Wooten--Winer--Weaire (WWW) bond-switching approach and with other recent approximants to ideal $a$-Si. More broadly, our work provides a platform for studies of band tails, optical properties, and transport in $a$-Si.
\end{abstract}

\maketitle

Amorphous silicon ($a$-Si) has long been thought of as a continuous random network (CRN) material \cite{Zachariasen-32-10}, with fourfold coordination ($N=4$ for all atoms) but without long-range structural order. 
The CRN model is an idealized picture that approximates the experimental reality, in which coordination defects ($N \neq 4$) are unavoidable \cite{Laaziri-99-11}, yet it is central to our fundamental understanding of disordered networks. 
Approaching the defect-free limit is relevant for precision instruments such as LIGO gravitational-wave detectors that rely on hydrogen-free $a$-Si for its low mechanical loss and correspondingly significant reduction in thermal noise \cite{Birney-18-11}.

Computationally, the CRN is often approximated using the Wooten--Winer--Weaire (WWW) bond-switching algorithm: starting from a crystalline ($c$-) Si structure, second-neighbor bonds are rearranged to change the network topology, whilst constraining it to $N=4$ throughout \cite{Wooten-85-04}. Other protocols do not enforce fourfold coordination, such that almost all reported defect-free $a$-Si models have been generated with the WWW protocol \cite{zongo_amorphous_2025}.

Despite the existence of several defect-free structural models, to the authors' knowledge, none has succeeded in demonstrating quantitative agreement with the experimental bandgap of $a$-Si, which varies between 1.5 and 1.8 eV, depending on the preparation method \cite{klazes_determination_1982}. This is often due to localized states at tails of the valence- and conduction-band edges, thus reducing the electronic gap. Previous studies have focused on the effect of hydrogenation on cleaning up the mid-gap states by saturating dangling bonds and reducing network strain, thereby modifying the band tails and opening up the bandgap \cite{Ching-80-03, Allan-98-03, meidanshahi_electronic_2019}.
An ideal, defect-free network would not require hydrogenation, instead presenting a `clean' electronic gap. 

We recently created a large dataset of $a$-Si structures from a series of melt-quench molecular-dynamics simulations \cite{Rosset-25-03}, driven by an efficient machine-learning-based interatomic potential (MLIP) \cite{Morrow-22-09, Si-GAP-18}, systematically sampling the configurational space of $a$-Si by varying the mass density and quench rate. From this library of thousands of disordered structures, we discovered a defect-free 216-atom model quenched at a rate of 10$^{10}$~K/s, denoted as `Rosset-10' in the following. We note that no coordination criterion was imposed during the simulation, so the system was not constrained to be defect-free, unlike in the WWW protocol. The generation of a defect-free $a$-Si network by melt-quenching is therefore a rare and stochastic event, and requires a large number of attempts, which can be accelerated using MLIPs \cite{Rosset-25-03, supplement}. 

\begingroup
\setlength{\tabcolsep}{4.5pt}
\renewcommand{\arraystretch}{1.15} 
\begin{table*}[ht]
    \centering
    \caption{Summary of the relaxed $a$-Si models discussed in the present study, detailing: generation protocol; number of atoms and percentage of 4-fold coordinated ones; mass density; porosity; percentage of locally crystal-like environments by Polyhedral Template Matching (PTM) and Common Neighbor Analysis (CNA); structural similarity to cubic diamond-type Si ($c$-Si) calculated using the SOAP kernel \cite{Bartok-13-05}; and energy relative to $c$-Si as predicted by HSE06-level DFT computations.}
    \label{tab:summary}
    \begin{tabular}{lccccccccc}
        \toprule\toprule
        Sample & Protocol &  \makecell{No.~of \\ atoms}  & \makecell{4-fold \\ coord.~(\%)} & \makecell{Density \\ (g/cm$^{3}$)}  & \makecell{Porosity \\ (\%)} &\makecell{PTM \\ (\%)} & \makecell{CNA \\ (\%)} & \makecell{SOAP \\ sim.}  & \makecell{$\Delta E$ \\ (eV/at.)} \\
        \midrule      
        \textbf{WWW-216} \cite{djordjevic_computer_1995} & WWW bond-switching \cite{Wooten-85-04} & 216 & 100 & 2.304 &  0.31 & 0 & 7.9 & 0.895 & 0.196 \\ \addlinespace
        \textbf{Pedersen-Y} \cite{Pedersen-17-06} & DFT relaxations & 215 & 98.1 & 2.267 &  0.22 & 0 & 0 & 0.911 & 0.189\\ \addlinespace
        \textbf{Zongo-R2} \cite{zongo_amorphous_2025} & ARTn runs \cite{Mousseau-12} & 216 & 100 & 2.256 &  0.24 & 0.5 & 19.9 & 0.919 & 0.160 \\ \addlinespace
        \midrule
        \textbf{Rosset-10} \cite{Rosset-25-03} & \makecell{Melt-quench, $10^{10}$ K/s rate} & 216 & 100 & 2.223 & 0.28 & 0 & 0 & 0.910 & 0.138 \\ \addlinespace
        \bottomrule\bottomrule
    \end{tabular}
\end{table*}
\endgroup

In this Letter, we characterize the electronic structure of our `Rosset-10' structural model of $a$-Si, and show that it faithfully reproduces the value of the electronic bandgap from experiment. We compare it to three existing high-quality models of $a$-Si (Table~\ref{tab:summary}):
(i) \textit{`WWW-216'}, obtained from the WWW protocol \cite{Wooten-85-04} paired with the Keating potential \cite{Keating-66-05}. Its electronic structure has been characterized previously \cite{djordjevic_computer_1995, pan_topological_2008};
(ii) \textit{`Pedersen-Y'}, obtained from a set of DFT annealing, relaxation, and rescaling procedures \cite{Pedersen-17-06}. While this structure is not entirely defect-free (Table~\ref{tab:summary}), it agrees very well with various experimental measurements of short- and medium-range structural order;
(iii) \textit{`Zongo-R2'}, generated by the Activation Relaxation Technique nouveau (ARTn), which moves the system along the potential-energy surface by searching for first-order saddle points, coupled with an MLIP \cite{zongo_amorphous_2025}. It is the only other defect-free 216-atom model that was not generated with the WWW protocol.
We note that Table~\ref{tab:summary} is not an exhaustive survey of simulation protocols for $a$-Si: others exist, such as Monte-Carlo \cite{Opletal-07-06} or hybrid approaches \cite{Cliffe-10-03, pandey_inversion_2016, Igram-18-07}.

We employed hybrid-level DFT computations, known to reproduce the experimental bandgap of $c$-Si \cite{Heyd-05-10, Xiang-13-03}, and previously used for $a$-Si \cite{Jarolimek-17-07, Morrow-24a}. Computations were carried out using a plane-wave energy cutoff of 540 eV and the projector augmented-wave formalism \cite{blochl_projector_1994}, as implemented in VASP \cite{kresse_efficient_1996, Kresse-99-01}. To evaluate all models on an equal footing, we first relaxed each model using the HSE06 functional \cite{Heyd-03-05, Heyd-06-06} at the $\Gamma$ point only. We then computed electronic properties using a $\Gamma$-centered $k$-point grid of 2$\times$2$\times$2 with the HSE06 functional. Further details are available in the Supplemental Material~\cite{supplement}. 

A common first criterion in validating a candidate amorphous structure is the predicted mass density. The density of $a$-Si thin films is estimated around 2.28 g/cm$^{3}$ from ion-implanted samples \cite{Custer-94-01, Laaziri-95-11, Laaziri-99-11}. The `Pedersen-Y' model is the closest to this value, likely since volume scaling was included in the generation protocol, but the other models fall within 1--3\% of the experimental value. The porosity is estimated using a fine 3D mesh and counting empty grid elements with a script developed in the context of Ref.~\citenum{Liu-26-02}. All models present low porosities, agreeing with reports that pores in $a$-Si are deposition artefacts  \cite{Moss-69-11} and should not occur in simulated quenched structures \cite{Lewis-22-03}.
The higher porosity of the `WWW-216' model arises from $m>8$ rings \cite{supplement}.

To ascertain the CRN nature of the models, we probe the existence of local ordering and compare three common indicators, viz.\ Polyhedral Template Matching (PTM) \cite{Larsen-16-05}, Common Neighbor Analysis (CNA) \cite{Honeycutt-87-09, Clarke-93-06, Stukowski-12-05, Maras-16-08}, and the averaged SOAP kernel similarity \cite{Bartok-13-05} to diamond-type (\textbf{dia}) Si, where 1 is identical. CNA appears more sensitive than PTM, indicating crystallinity in both the `WWW-216' and `Zongo-R2' models, while PTM only detects ordering in the latter. This crystallinity is consistent with reports of ordering in the original study \cite{zongo_amorphous_2025}. In SOAP terms, the `Zongo-R2' model is closest to \textbf{dia}, whereas `WWW-216' is most dissimilar. `Rosset-10' shows no apparent crystallinity and an average SOAP similarity to \textbf{dia} relative to the other models and the literature \cite{Bernstein-19, Rosset-25-03}. Beyond global structural characteristics, the Supplemental Material includes metrics of short- and medium-range order \cite{supplement}.

The four models show large variations in their energies relative to \textbf{dia}, as computed with HSE06. In particular, the `Rosset-10' model has the lowest energy among the four. We have previously shown that paracrystalline models have lower energies on average than fully disordered networks \cite{Rosset-25-03}, hence this lower energy could reflect crystalline ordering that was not captured by PTM, CNA or SOAP. Yet, the `Rosset-10' model provides excellent agreement with the calorimetrically-determined enthalpy of crystallization of $\Delta H$ =  $0.142 \pm 0.003$ eV/atom obtained from ion-implanted samples \cite{Roorda-12-12}. Importantly, this means that only the `Rosset-10' model fulfills the energy criterion set out in Ref.~\citenum{Drabold-11}, that requires the energy of a candidate ideal $a$-Si model to fall within the range of 0.07 to 0.15 eV/atom above that of {\bf{dia}}. 

\begin{figure*}[ht]
    \centering
    \includegraphics[]{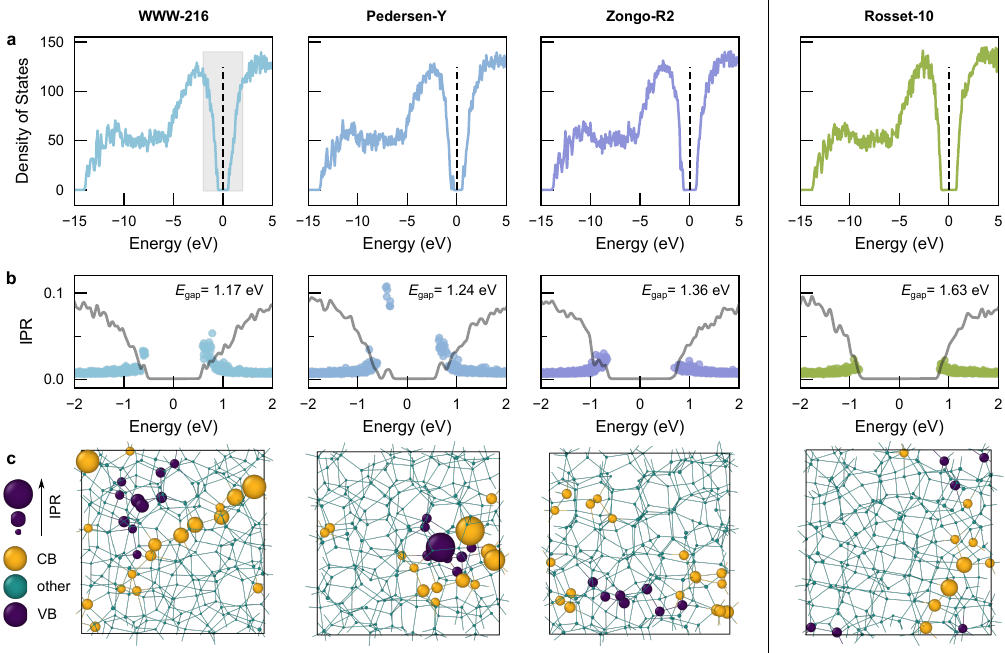}
    \caption{Electronic structure. (a) Electronic densities of states of the four structural model considered in this study, where a dashed vertical line indicates the Fermi level ($E_\text{F}$)~defined as the middle of the gap and centered at 0 eV. Enlarged plots are available in the Supplemental Material \cite{supplement}. (b) Inverse Participation Ratio (IPR) plots in the region of the bandgap for each model, as shaded in gray for the `WWW-216' model. The IPR is evaluated for each band and $k$-point pair; thick gray line overlays show the densities of states. The quoted $E_{\text{gap}}$~value refers to the energy difference between the lowest occupied conduction-band (LUMO) state and the highest occupied valence-band (HOMO) state, omitting the mid-gap states for the `Pedersen-Y' model. (c) Visualizations of the spatial localization of the highest IPR sites for each structure using OVITO \cite{ovito}, where atoms are color-coded by their IPR contribution to the VB, CB, or other as purple, yellow, or teal, respectively. The atomic radii are scaled proportionally to the relative IPR magnitudes, normalized against the largest IPR contribution across all structural models.}
    \label{fig1}
\end{figure*}

We now turn to the electronic densities of states (DOS), plotted in Fig.~\ref{fig1}(a). 
The shape of the DOS is well reproduced throughout, with the 3$s$ orbitals mainly contributing in the $-15$ to $-5$ eV range, and the 3$p$ orbitals between $-5$ and 0 eV. The relative heights of the valence-band (VB) and conduction-band (CB) maxima are similar and agree well with previous literature \cite{Thorpe-71-12, Dong-98-03}. All three defect-free structures have clean bandgaps free of mid-gap states, while the DOS of the `Pedersen-Y' model does have such states, associated with the aforementioned coordination defects (Table \ref{tab:summary}).

Other important features of the DOS are the exponential tails at the VB and CB edges, known as Urbach tails \cite{Urbach-53-12}, which have been resolved experimentally \cite{Aljishi-90-06}. These tails have been linked to atomic filaments, whose electron eigenstates form interconnected clusters of charge \cite{Dong-98-03}, and also to topological filaments associated with bond stretching and bending \cite{pan_topological_2008, Drabold-11-01}. The `WWW-216' and `Zongo-R2' models show large tails at their CB and VB edges, respectively, while the `Rosset-10' model shows small and narrower VB and CB tails. 

\begin{figure*}[ht]
    \centering
    \includegraphics[]{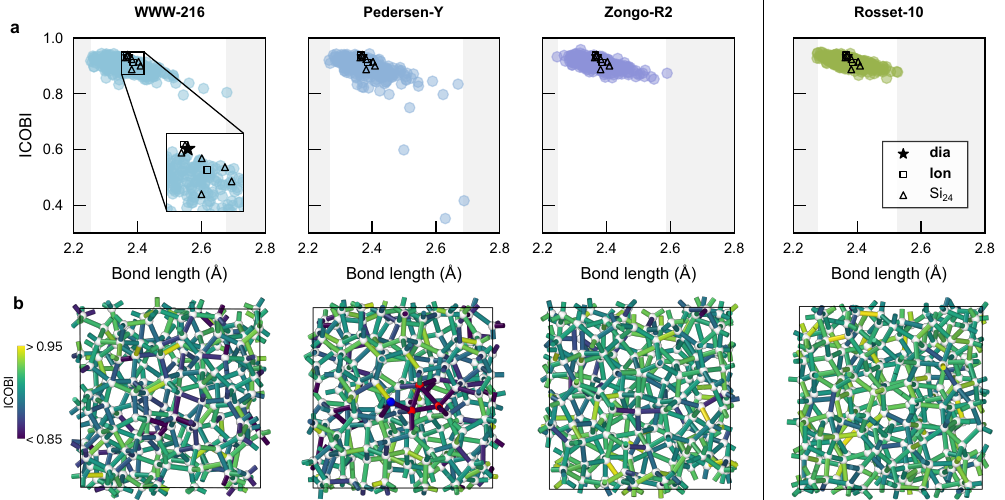}
    \caption{Chemical bonding. (a) Scatter plots of integrated pairwise crystal orbital bond index (ICOBI$^{(2)}$) values against bond lengths, where gray shading shows the edges of the bond-length distributions. An inset highlights relevant $c$-Si modifications: cubic diamond-type (``{\bf dia}''; star), hexagonal diamond-type (``{\bf lon}''; square) and the open-framework structure of Ref.~\citenum{Kim-15-02} (``Si$_{24}$''; triangle). (b) Structure visualizations of the models using OVITO \cite{ovito}, where atoms are color-coded by coordination (blue/white/red for 3/4/5-fold coordination, respectively), and bonds are color-coded by ICOBI value.}
    \label{fig2}
\end{figure*}

We report the estimated values of the electronic bandgap, $E_{\text{gap}}$, in Fig.~\ref{fig1}(b), defined as the difference in energy between the lowest occupied CB (LUMO) state and the highest occupied VB (HOMO) state. For the `Pedersen-Y' model, we exclude the localized mid-gap states when defining the bandgap.
There is a large variation in the estimated $E_{\text{gap}}$ values between models, directly reflecting the differences in the Urbach tails. Previously reported values for $a$-Si in the computational literature range between 0.8 and 1.35 eV \cite{Durandurdu-00-12, Barkema-00-08}; the `WWW-216', `Pedersen-Y' and `Zongo-R2' models are consistent with these. The `Rosset-10' model has a substantially larger $E_{\text{gap}}$, which provides excellent and unprecedented agreement with the experimentally reported range of optical bandgaps of $a$-Si, typically between 1.5 and 1.8 eV \cite{Thutupalli-77-02, klazes_determination_1982, Park-01-02, Huang-21-01, supplement}.
We note that our results are obtained from computations with a $2\times2\times2$ $k$-point grid, and that the estimated $E_\text{gap}$ varies with $k$-sampling and the associated dispersion in the supercell \cite{Drabold-90-09, Prasai-16-06}. Identical calculations at the $\Gamma$-point result in gaps that are systematically larger, by 0.02 to 0.12~eV, giving $E_{\text{gap}}=1.71$~eV for `Rosset-10'.

To probe these variations in the electronic structure, we focus on the spatial localization of states around the Fermi level, $E_{\text{F}}$, in Fig.~\ref{fig1}(b), plotting the DOS together with inverse participation ratio (IPR) values \cite{Thouless-74-10}: 
\[\text{IPR}_{nk}=\dfrac{\Sigma_a|c_{nka}|^4}{(\Sigma_a|c_{nka}|^2)^2}\]
where $n$ is the band index, $k$ is the $k$-point index, $a$ is the atomic orbital and $c$ is the wavefunction coefficient, as calculated with \textsc{PyProcar} \cite{Herath-20-10, Lang-24-09}. The IPR measures how localized or extended  the electron states are \cite{Dong-98-03, Durandurdu-00-12}: low values ($\rightarrow 0$) correspond to delocalized or extended states that are distributed across the structure, while high values ($\rightarrow 1$) correspond to spatially localized states. However, it should be noted that it is not possible to distinguish unambiguously between delocalized and weakly localized states when the localization length is comparable to the simulation-box size, here $\approx15$~Å.

The `Pedersen-Y' model has highly localized states adjacent to the VB edge, arising from the dangling-bond defect. It also displays some localization at the CB edge, associated with a broader conduction-band tail. The `WWW-216' and `Zongo-R2' models reveal appreciable localization at the CB edge and the VB edge, respectively. The `Rosset-10' model exhibits no significant electron-localization, instead seemingly having fully extended (delocalized) states throughout both the VB and CB edges.
It appears that the degree of localization around the bandgap is inversely correlated with the width of the gap: the models with the smallest $E_{\text{gap}}$ values, i.e., `WWW-216' and `Pedersen-Y', show the most tail localization, while the `Rosset-10' model shows no localization and has the largest bandgap.

The width of the bandgap and the degree of localization are associated with the degree of structural disorder and the presence of coordination defects, respectively \cite{Pan-08-05}. To examine the spatial character of the localized states, Fig.~\ref{fig1}(c) shows the atomic regions contributing most strongly to the IPR for states near the band edges, where the atomic radii are scaled proportionally to the largest IPR contribution in the structure. The `WWW-216' and `Pedersen-Y' models exhibit localization on relatively small groups of atoms, arranged in compact clusters. This is consistent with the larger IPR values observed for these models, as higher IPR indicates stronger confinement of the electronic state to a small number of sites \cite{Pan-08-05}. By contrast, the `Zongo-R2' and `Rosset-10' models show lower IPR values at the band edges and correspondingly more spatially extended localization patterns, which appear as interconnected chains within the network. This behavior is consistent with earlier studies showing that band-tail states in $a$-Si arise from chains of distorted bonds rather than isolated point-like defects \cite{Fedders-98-12, Pan-08-05, Inam-10, Drabold-11-01}. In particular, mildly short bonds contribute predominantly to VB tail states, while mildly long bonds contribute predominantly to CB tail states, a trend that is reproduced by the `Pedersen-Y' and `Rosset-10' models, as shown in the Supplemental Material \cite{supplement}.

We turn to the chemical bonding in these structures to understand the large variation in $E_{\text{gap}}$ and the exceptionally extended states in the `Rosset-10' model. In Fig.~\ref{fig2}(a), we plot the integrated crystal orbital bond index (ICOBI$^{(2)}$), obtained with LOBSTER \cite{Maintz-13, Nelson-20, muller_crystal_2021, supplement}. ICOBI$^{(2)}$ quantifies the pairwise bond order, such that strong covalent bonds have a value close to 1 \cite{muller_crystal_2021}.
Stretched bonds in the `WWW-216' and `Pedersen-Y' models result in lower ICOBI values, but there are strong outliers in `Pedersen-Y' falling beyond the exponential bond-length--bond-strength relationship \cite{Pauling-47-03}. The distributions for the `Zongo-R2' and `Rosset-10' models are qualitatively similar, but the bond-length distribution of the latter is narrower, resulting in a more compact scatter.
For comparison, an inset shows ICOBI values for crystalline allotropes of Si: cubic and hexagonal diamond-type, and an open-framework structure \cite{Kim-15-02}.

To resolve the spatial character of bonding, we visualize the structural models in Fig.~\ref{fig2}(b), coloring bonds by their ICOBI values.
The low values in the `Pedersen-Y' model arise from a 3-membered ring formed by the overcoordinated Si atoms (red). The three long bonds strain their local environment into a larger cluster with disrupted bonding, in turn leading to the localized states in the gap and in the band tails. The bonds around the undercoordinated Si atom (blue) also leads to local strain and higher ICOBI values. In the `WWW-216' model, some bonds with low ICOBI values belong to 4-membered rings, while others form linear chains, or isolated stretched bonds.
Conversely, the ICOBI values are more uniformly distributed in the `Zongo-R2' and `Rosset-10' models, and the bonds with low ICOBI values do not cluster.

A model of $a$-Si with no localized states, a narrow bond-angle distribution, and low excess energy with respect to \textbf{dia} has been hypothesized to represent an \textit{ideal} $a$-Si network \cite{Durandurdu-00-12}. All three criteria are fulfilled with the `Rosset-10' model, as shown in Fig.~\ref{fig1}, Fig.~S2 \cite{supplement}, and Table~\ref{tab:summary}. We further propose that an `ideal' $a$-Si model must have a narrow distribution of bond lengths without excessively short nor long bonds, a criterion that appears necessary to achieve an absence of localized states at the VB and CB band edges, respectively. 

Looking forward, our model could serve as a platform to further understand the effect of defects and dopants in $a$-Si \cite{Winer-88-06}, or to make accurate predictions of carrier mobility \cite{Lee-22-02}. These effects are of interest since $a$-Si is used in both intrinsic and doped forms in solar-cell heterojunctions \cite{Lin-23-08}. 
The model's large bandgap may be of interest in studying the optical absorption of $a$-Si, relevant for LIGO applications \cite{Birney-18-11}.
Our model could also be used in the future to investigate why $a$-Si has a larger bandgap than the indirect gap in \textbf{dia}, in an analogous study to the one recently conducted on $c$-Si in Ref.~\citenum{Oliphant-25}.
More broadly, this model of an ideal random network opens up new avenues for the exploration of the electronic structure of CRN-like disordered materials.

\begin{acknowledgments}
This work was supported by UK Research and Innovation [grant number EP/X016188/1]. 
C.U. acknowledges support from the  laboratory directed Research and Development (LDRD) program at Los Alamos National Laboratory (LANL) through a Director’s Postdoctoral Fellowship (Project No. 20240877PRD4).
S.R.E. is grateful to the Leverhulme Trust (UK) for a Fellowship.  
D.A.D. acknowledges support from the U.S. National Science Foundation (Project No. MRI-2320493). 
We are grateful for computational support from the UK national high performance computing service, ARCHER2, for which access was obtained via the UKCP consortium and funded by EPSRC [grant number EP/X035891/1].
Data supporting this work are provided at https://github.com/vldgroup/papers-defect-free-a-si.
\end{acknowledgments}

\end{document}


\title{\Large {\bf Supplemental Material for}\\ {`A Defect-Free Model of Amorphous Silicon with Pristine Electronic Structure'}}

\author[1]{Louise~A.~M.~Rosset}
\author[2]{Chinonso~Ugwumadu}
\author[3]{Stephen~R.~Elliott}
\author[4]{David~A.~Drabold}
\author[1]{Volker~L.~Deringer\thanks{volker.deringer@chem.ox.ac.uk}}

\affil[1]{Inorganic Chemistry Laboratory, Department of Chemistry, University of Oxford, Oxford, UK}
\affil[2]{Quantum \& Condensed Matter (T-4) Group, Los Alamos National Laboratory, Los Alamos, NM, USA}
\affil[3]{Physical and Theoretical Chemistry Laboratory, Department of Chemistry, University of Oxford, Oxford, UK}
\affil[4]{Department of Physics and Astronomy, Ohio University, Athens, OH, USA}

\date{}

\maketitle

\setstretch{1.5}
\setcounter{suppfigure}{1}

\newpage

\section{Methods}

\subsection{Structural models}
The `Rosset-10' model originates from a dataset generated in Ref.~\citenum{Rosset-25-03}, comprised of structures of 64, 216, 512, and 1,000 atoms. None of the structures of 512 or 1,000 atoms were defect-free. The structures of 64 atoms were omitted due to their very small size.
Within the subset of 782 structural models of 216 atoms, only four were defect-free. From these four, two showed substantial paracrystalline content according to CNA and PTM analysis (see main text).
This would give a rate of generating continuous random networks that are defect-free of $\approx 0.26\% $. 
However, if we consider that only 381 structures were generated with a slow quench rate, that is, either $10^{10}$or $10^{11}$~K/s, then this rate increases to  $\approx 0.52\% $.

Of the two defect-free CRN-like models, we focused on the one with a larger bandgap and narrower bond-length distribution, denoted `Rosset-10' as it was generated with a quench rate of $10^{10}$~K/s. The other model, generated with a quench rate of $10^{11}$~K/s, showed a smaller bandgap of 1.42~eV, with localization on the edge of the valence band. It has a wider bond-length distribution, with $\approx$ 0.4$\%$ of bonds with $l>2.6$~Å.

\subsection{DFT computations}
DFT computations were carried out in VASP v6.5.1, \cite{kresse_ab_1994, kresse_efficiency_1996} using the projector augmented-wave formalism. \cite{blochl_projector_1994, kresse_efficient_1996}
All structural models were relaxed with HSE06 using a plane-wave energy cutoff of 540 eV at the $\Gamma$ point, with a convergence criterion of 10$^{-6}$ eV.

To accelerate our computations with the HSE06 functional, a pre-convergence with PBE was performed to a convergence criterion of 10$^{-6}$ eV. The WAVECAR and CHGCAR were then used to initialize the HSE06 self-consistent field (SCF) computation. 
The HSE computation was performed with a range-separation parameter of 0.2 Å$^{-1}$, a fraction of exact exchange of 0.25, a plane wave energy cutoff of 540 eV, a Gaussian smearing width of 0.05 eV, a $\Gamma$-centered $k$-point grid of 2$\times$2$\times$2, and a convergence criterion of 10$^{-6}$ eV/cell. 

We did not include non-spherical contributions related to the gradient of the density in the PAW spheres (\texttt{LASPH} tag). A test in which we set \texttt{LASPH=.TRUE.} for the HSE06 computation resulted in a change of 0.1~meV in the predicted $E_\text{gap}$ of the `Rosset-10' model, much smaller than the precision to which we report $E_\text{gap}$ (0.01~eV or 10 meV).

We used LOBSTER v5.1.1 for chemical-bonding analysis on the HSE06 SCF outputs, using 3s 3p basis functions, with a Gaussian smearing width of 0.05 eV. We consider chemical bonding interactions for pairs within a cutoff of 2.85 Å, and an energetic window from --15 eV to 10 eV relative to the Fermi level.

\subsection{Experimental bandgaps}

Table \ref{tab:bandgaps} summarizes experimental literature relating to measurements of the bandgap of amorphous silicon. 

\begingroup
\setlength{\tabcolsep}{4.5pt}
\renewcommand{\arraystretch}{1.15} 
\begin{table*}[h]
\refstepcounter{supptable}
\centering
\begin{threeparttable}
    \footnotesize
    \caption{Summary of reported bandgaps from experimental literature, detailing the sample preparation protocol, the sample type, the bandgap measurement method, the bandgap ($E_\textrm{gap}$), and the year and number of the literature reference.}
    \label{tab:bandgaps}
    \begin{tabular}{lccccc}
        \toprule\toprule
        Protocol & Sample type & Method & $E_\textrm{gap}$ (eV)& Year & Reference \\
        \midrule      
        Evaporation & Thin film & Tauc with $r = 2$ & 1.82--1.83 & 1977 & \citenum{Thutupalli-77-02} \\ 
        Evaporation & Thin film & Tauc with $r = 2$  & 1.5 & 1993 & \citenum{Carlsson-93-07} \\
        Sol--gel synthesis & Nanoparticle & Photoluminescence & 1.53 -- 1.82 & 2021 &\citenum{Huang-21-01} \\
        Sputtering & Thin film & Tauc with $r = 3$  & 1.1--1.4 & 1982 &\citenum{klazes_determination_1982} \\ 
        PECVD\tnote{1} & Thin film & Tauc with $r = 3$ & 1.6--1.8 & 1982 &\citenum{klazes_determination_1982} \\ 
        PECVD\tnote{2} & Thin film & Tauc with $r = 2$ & 1.49 & 1993 &\citenum{Berntsen-93-11} \\
        PECVD\tnote{3} & Quantum dot & Photoluminescence & 1.56 & 2001 &\citenum{Park-01-02}\\ 
        \bottomrule\bottomrule
    \end{tabular}
    \begin{tablenotes}
    \item[1] Hydrogen content in these samples is not discussed in the reference work.
    \item[2] Samples contain less than 0.3 at.-\% H.
    \item[3] Samples were broadly described as having `very little' hydrogen.\cite{Park-01-02}
    \end{tablenotes}
\end{threeparttable}
\end{table*}
\endgroup

The entries in Table \ref{tab:bandgaps} vary in preparation protocol, sample type, and in the approximation used to calculate the $E_\textrm{gap}$ value from experimental findings.
Broadly, the reported bandgaps range from 1.5 to 1.8~eV. We omit the sputtered samples from Ref.~\citenum{klazes_determination_1982} from this range as they appear untypically low compared to other reports for $a$-Si, and as the original paper reports that sputtering has yielded a `different material' to the samples obtained in the same study from glow discharge deposition\cite{klazes_determination_1982} (denoted `PECVD', plasma enhanced chemical vapor deposition, in Table \ref{tab:bandgaps}).

We note that samples synthesized by PECVD from \ce{SiH4} or related precursors may contain some amount of hydrogen, which may artificially increase the reported bandgap value. Hydrogen contents for each of these samples are discussed as footnotes in Table \ref{tab:bandgaps}.

We also note that most of these experiments were performed over 30 years ago, and that new, high quality, experimental data would substantially strengthen the comparison between experimental results and computational models.

\subsection{Density of states}

\begin{figure*}[h]
     \centering
     \includegraphics[]{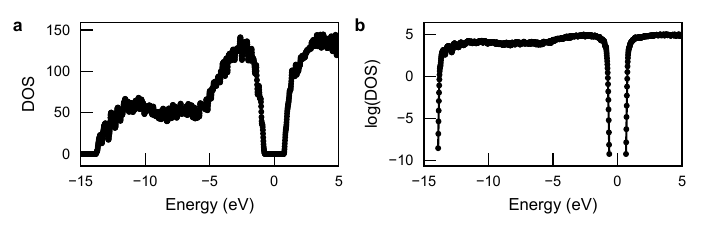}
     \caption{Plots of (a) the density of states and (b) the logarithm of the density of states for the `Rosset-10' model.}
     \label{dos-res}
     \refstepcounter{suppfigure}
 \end{figure*}

We assess whether the resolution of the density of states computation (Fig.~\ref{dos-res}(a)) is sufficient by plotting the $\log$ of the density of states (Fig.~\ref{dos-res}(b)), here shown for the `Rosset-10' structure as an example. The smooth behavior captured in the logarithmic plot suggests that the resolution is sufficient.

\subsection{Molecular-dynamics annealing}
We performed a 1 ns annealing run at 300 K and 1 bar in the \(NpT\) ensemble using the Si-GAP-18 model \cite{Bartok-18-12} to obtain well-equilibrated structural models. The Si-GAP-18 model has been extensively validated against experimental data for $a$-Si \cite{Deringer-18-06, Deringer-21-01}. The last 100 ps were used for structural analysis. In Table~\ref{tab:MD}, we investigate the effect of the annealing treatment on some global characteristics, comparing to the structures before annealing (Table~1).

\begin{table}[h]
\refstepcounter{supptable}
     \centering
     \caption{Global characteristics of the HSE06-relaxed structural models (`Static') compared to the MD-annealed models (`MD'): mass density, coordination, percentage of locally crystalline-like environments by PTM and CNA, SOAP similarity to \textbf{dia}, and the Si-GAP-18 predicted energy of the model relative to \textbf{dia}. Results for the MD-annealed models are obtained from averaging over the last 100 ps.}
    \resizebox{\linewidth}{!}{
    \begin{tabular}{cccccccccccccc}
    \toprule\toprule
         \multirow{ 2}{*}{Sample} & \multicolumn{2}{c}{$\rho$ (g/cm$^3$)} & \multicolumn{2}{c}{4-fold coord.} & \multicolumn{2}{c}{PTM (\%)} & \multicolumn{2}{c}{CNA (\%)} & \multicolumn{2}{c}{SOAP sim. \textbf{dia}} & \multicolumn{2}{c}{$\Delta E$ (eV/at.)} \\
          & Static & MD & Static & MD & Static & MD & Static & MD & Static & MD & Static & MD \\
          \midrule
          WWW-216 & 2.304 & 2.251 & 100 & 98.9 & 0 & 0 & 7.9 & 3.3 & 0.895 & 0.890 & 0.193 & 0.226 \\
          Pedersen-Y & 2.267 & 2.276 & 98.1 & 97.4 & 0 & 0 & 0 & 4.9 & 0.911 & 0.904  & 0.186 & 0.217 \\
          Zongo-R2 & 2.256 & 2.260 & 100 & 99.9 & 0.5 & 0.16 & 19.9 & 15.7 & 0.919 & 0.912  & 0.155 & 0.191 \\
          Rosset-10 & 2.223 & 2.239 & 100 & 100 & 0 & 0 & 0 & 3.1 & 0.910 & 0.904 & 0.138 & 0.175 \\
    \bottomrule\bottomrule
    \end{tabular}
    }
\label{tab:MD}
\end{table}

After a long 1~ns annealing run, both the `Zongo-R2' and `Rosset-10' models retained ideal fourfold connectivity (to within $\approx 0.1$\%), whereas the other two models contained a small number of coordination defects.

\clearpage

\section{Supplemental results}
\subsection{Structural analysis}

\begin{figure}[h]
    \centering
    \includegraphics[width=\linewidth]{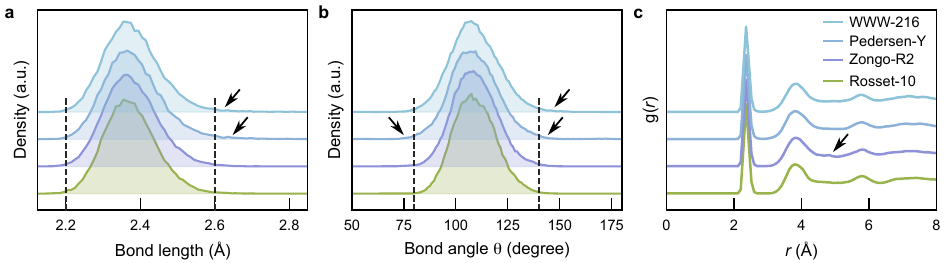}
    \caption{Characteristics of short- and medium-range order. (a) Bond-length distribution, where arrows highlight the distribution tails, and dashed vertical lines indicate short ($r<2.2$Å) and long ($r>2.6$Å) bonds. (b) Bond-angle distribution, where dashed vertical lines indicate small ($\theta<80\degree$) and large ($\theta>140\degree$) angles, and arrows highlight the distribution tails. (c) Radial distribution function, $g(r)$, where an arrow indicates an enhancement in the `Zongo-R2' model. Distributions are vertically offset for clarity, and averaged over the last 100 ps of a 1 ns $NpT$ annealing simulation at 300 K and 1 bar, driven by the Si-GAP-18 potential.\cite{Bartok-18-12}}
    \label{sro}
    \refstepcounter{suppfigure}
\end{figure}

We report metrics of short- and medium range order in Fig.~\ref{sro}, first plotting bond-length distributions. All four structural models show maxima around 2.38~Å (Table~\ref{tab:sro}), which compares well to the experimental average of 2.35~Å (ion-implanted samples; Ref.~\citenum{Laaziri-99-11}). An experimental standard deviation of $\approx$ 0.065~Å was reported, wider than that of the models obtained by peak fitting ($\approx 0.04$~Å), but the authors of Ref.~\citenum{Laaziri-99-11} warn that some broadening may arise from vibrational disorder and insufficient experimental resolution. The tails of the bond-length distributions are where the models differ structurally: while `Rosset-10' and `Zongo-R2' show fewer than 0.5$\%$ of short bonds with $l<2.2$~Å or of long bonds with $l>2.6$~Å, the `WWW-216' and `Pedersen-Y' models show significant tails from overstretched bonds, 1.0$\%$ and 1.4$\%$ respectively, as indicated by arrows in Fig.~\ref{sro}. We note that the bond-length distribution presented in Fig.~\ref{sro}(a) is broader than the one shown in Fig.~3 of the main text -- Fig.~\ref{sro}(a) shows the distributions obtained from a molecular-dynamics annealing simulation, while the distribution in Fig.~3 is the static HSE06-relaxed structure. The annealing simulations induce thermal vibrations which broaden the distribution of bond lengths.

We plot the bond-angle distributions in Fig.~\ref{sro}(b), with complementary information detailed in Table~\ref{tab:sro}. All four models agree reasonably with the experimental average of 107.83\textdegree $\pm$ 0.97\textdegree ~(Ref.~\citenum{Laaziri-99-11}). The `Zongo-R2' model has a slightly narrower bond-angle distribution, likely due to paracrystalline ordering; `WWW-216' and `Pedersen-Y' show broader tails, with distorted small ($\theta<80$\textdegree) and large angles ($\theta>140$\textdegree).

The radial distribution functions [Fig.~\ref{sro}(c)] are similar across all four models, with well-defined peaks for the first- and second-neighbor shells. The `Zongo-R2' model shows a small feature around 4.5 Å: an indicator of paracrystalline ordering observed in experimental \cite{Fortner-89-03, Laaziri-99-04} and computational studies,\cite{Nakhmanson-01-05, Rosset-25-03} consistent with metrics of crystallinity given in Table~1 of the main text. The `Rosset-10' model does not show such a feature, supporting its CRN character.

\begingroup
\setlength{\tabcolsep}{4.5pt}
\renewcommand{\arraystretch}{1.15} 
\begin{table*}[h]
\refstepcounter{supptable}
    \centering
    \footnotesize
    \caption{Mean and standard deviation of the bond-length and bond-angle distributions for each structural model.}
    \label{tab:sro}
    \begin{tabular}{lcccc}
        \toprule\toprule
        Sample & 〈$r$〉~(Å) & $\sigma_r$~(Å) &〈$\theta$〉~(\textdegree) & $\sigma_\theta$~(\textdegree) \\
        \midrule      
        \textbf{WWW-216} & 2.382 & 0.041 & 109.2 & 10.48 \\ \addlinespace
        \textbf{Pedersen-Y} & 2.379 & 0.039 & 108.9 & 11.03 \\ \addlinespace
        \textbf{Zongo-R2} & 2.375 & 0.046 & 109.2 & 8.84 \\ \addlinespace
        \midrule
        \textbf{Rosset-10} & 2.376 & 0.044 & 109.3 & 10.05 \\ \addlinespace
        \bottomrule\bottomrule
    \end{tabular}
\end{table*}
\endgroup

\begin{figure*}[t]
     \centering
     \includegraphics[]{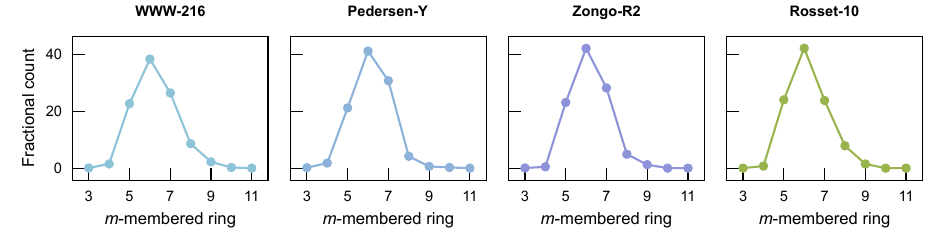}
     \caption{Distributions of the shortest closed-path rings, calculated with \textsc{matscipy} \cite{Grigorev2024} obtained from averaging over the last 100 ps of a 1 ns molecular-dynamics simulation at 300~K.}
     \label{rings}
     \refstepcounter{suppfigure}
 \end{figure*}

We plot the ring-size distributions in Fig.~\ref{rings}. The `WWW-216' model has a slightly lower count of $m=6$ membered rings than the other three structures, and instead has a large count of $m>8$ rings. Unlike the other three models, the `Rosset-10' has a higher count of $m=5$ membered rings than $m=7$ rings.

\begin{figure*}[t]
     \centering
     \includegraphics[]{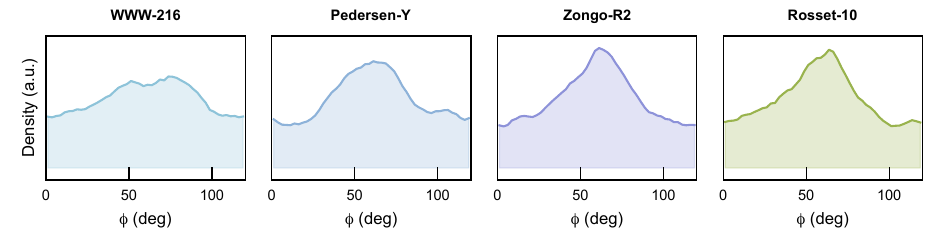}
     \caption{Dihedral-angle distributions, obtained from averaging over the last 100 ps of a 1 ns molecular-dynamics simulation at 300~K.}
     \label{dihedrals}
     \refstepcounter{suppfigure}
 \end{figure*}

We show the distributions of dihedral angles in Fig.~\ref{dihedrals}. The dihedral-angle distribution has been linked to the existence of an enhancement in the RDF around $\approx4.5$Å,\cite{Laaziri-99-11} and discussed as an indicator of paracrystallinity.\cite{Rosset-25-03} The `Zongo-R2' and `Rosset-10' models have similarly sharp distributions, which may be an indicator of paracrystalline ordering in the latter which is not apparent from the other structural metrics presented here, namely PTM, CNA, SOAP, and the RDF.

\begin{figure*}[t]
     \centering
     \includegraphics[]{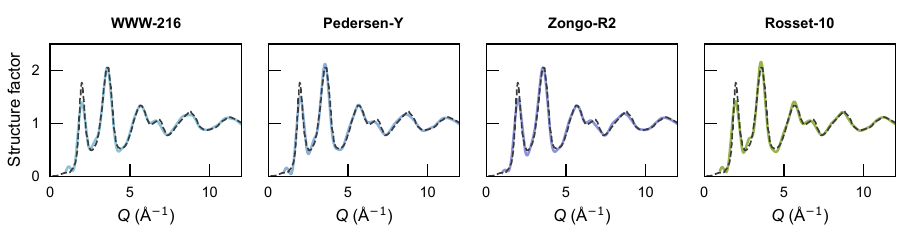}
     \caption{Computed structure factor of each structural model, obtained from the Fourier transform of the radial distribution function from OVITO \cite{ovito}. Data are averaged over the last 100 ps of a 1 ns MD simulation at 300~K. Experimental data from Ref.~\citenum{Laaziri-99-11} are plotted as a dashed black line.}
     \label{sq}
     \refstepcounter{suppfigure}
 \end{figure*}

We also plot the structure factors of the models in Fig.~\ref{sq}, comparing to the high-quality experimental data from Ref.~\citenum{Laaziri-99-11}. We note that significant size effects must be considered for models of only 216 atoms. The `Rosset-10' model provides the best match to the high-$Q$ part of the structure factor, but presents a shoulder at the foot of the second sharp diffraction peak. Only the `Zongo-R2' model appears completely devoid of this feature.

\begin{figure*}[t]
     \centering
     \includegraphics[width=\textwidth]{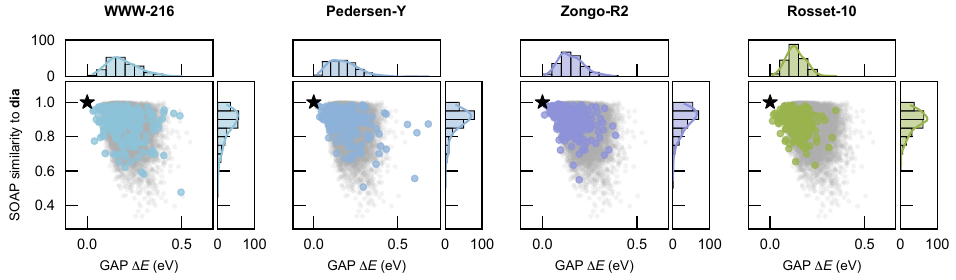}
     \caption{Structure and energetics of individual atomic environments, plotted in the style of Fig.~2 from Ref.~\citenum{Rosset-25-03}. Scatter plot of the GAP--predicted atomic energies relative to cubic-diamond-type Si ({\bf dia}), against the atomistic SOAP similarity to {\bf dia} for each structural model. A black star indicates the ideal {\bf dia} environment. An underlying set of data points (gray) shows the distribution of local atomic environments in the dataset of disordered $a$-Si structures of 216 atoms from Ref.~\citenum{Rosset-25-03}. Histograms of the total distribution and kernel density estimates are shown for each axis.}
     \label{SOAP_deltaE}
     \refstepcounter{suppfigure}
 \end{figure*}

Finally, we probe the information provided by local environments, which has been shown to be valuable in quantifying local disorder.\cite{Bernstein-19, Rosset-25-03} Here, we plot the local SOAP similarity to \textbf{dia} of each atomic environment against its atomic energy above that of \textbf{dia} in Fig.~\ref{SOAP_deltaE}. 
To compare the local environments in these models to those from well-annealed disordered samples, we take the subset of disordered `CRN'-like networks of 216 atoms generated in Ref.~\citenum{Rosset-25-03}. We show all the local environments from this subset in gray. As follows from the chemical-bonding analysis in Fig.~3, the `Rosset-10' model shows a tighter scatter and more homogeneous disordered environments than the other models, both in terms of local structure and local energy.

We calculate the information entropy of these distribution using: \(H=-\sum_ip_i\times\log{p_i}\), where $p_i$ is the probability of bin $i$. 
In relative terms, low entropy is indicative of a more `predictable' distribution, such that the probabilities are concentrated on a small number of bins; high entropy is indicative of a less `predictable' distribution where the probabilities are spread out across many different bins. The SOAP similarity to \textbf{dia} distributions have entropy values of 1.84 for `WWW-216', 1.68 for `Pedersen-Y', 1.67 for `Zongo-R2', and 1.64 for `Rosset-10'. The $\Delta E$ distributions have entropy values of 1.92 for `WWW-216', 1.90 for `Pedersen-Y', 1.71 for `Zongo-R2', and 1.49 for `Rosset-10'.

\subsection{Density of states}
\begin{figure*}[h]
     \centering
     \includegraphics[width=\linewidth]{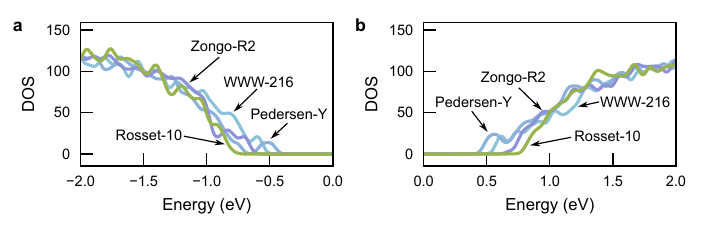}
     \caption{Enlarged plots of the density of states at the: (a) valence-; and (b) conduction-band edges for all four structural models.}
     \label{dos-tails}
     \refstepcounter{suppfigure}
 \end{figure*}

To further support the discussion of the Urbach tails from Fig.~1 of the main text, we show an enlarged plot of the valence- and conduction-band edges in Fig.~\ref{dos-tails}. The features are somewhat sharp from the low Gaussian broadening of 0.05 eV used in the DFT computations.

The $E_\text{gap}$ values in Fig.~1 are computed with a $k$-point grid of $2\times2\times2$. For a $\Gamma$-point only calculation, the computed $E_\text{gap}$ values are: 1.19 eV for `WWW-216', 1.05 eV for `Pedersen-Y', 1.50 eV for `Zongo-R2', and 1.71 eV for `Rosset-10'.

\subsection{Inverse Participation Ratio (IPR)}

\begin{figure*}[h]
     \centering
     \includegraphics[]{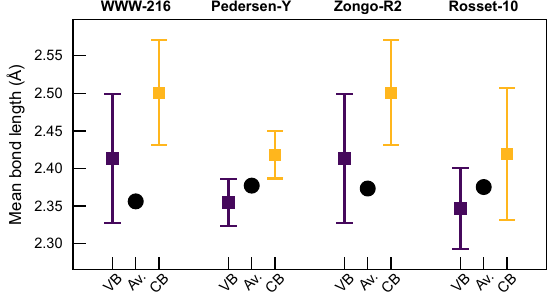}
     \caption{Mean bond length for atoms with a high Inverse Participation Ratio (IPR) at the edge of the valence (VB) and conduction bands (CB) of the four structural models. The mean overall bond length in the model is shown in black. Whiskers show the standard deviation. }
     \label{ipr_r}
     \refstepcounter{suppfigure}
 \end{figure*}

We examine the mean bond length of atoms contributing most strongly to the high-IPR states near the valence- and conduction-band edges in the four structural models in Fig.~\ref{ipr_r}. The resulting trends are consistent with earlier theoretical work showing that band-tail states in $a$-Si arise from correlated structural disorder rather than isolated point defects. Specifically, valence-tail states are associated with relatively short-bond networks, whereas conduction-tail states are associated with relatively long-bond networks.\cite{Fedders-98-12, Pan-08-05, Drabold-11-01} This trend is reproduced by the `Pedersen-Y' and `Rosset-10' models. For the `WWW-216' and `Zongo-R2' models, the contributions to both the VB and CB edges arise from longer-than-average bonds, with bonds contributing to the CB being longer on average than those contributing to the VB.

\subsection{Chemical bonding}

\begin{figure*}[h]
     \centering
     \includegraphics[width=\linewidth]{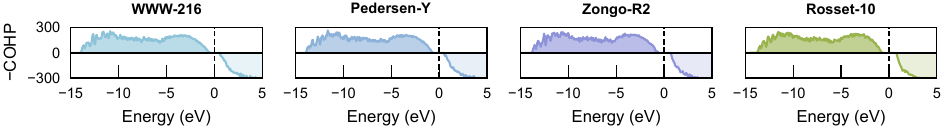}
     \caption{Negative crystal-orbital Hamilton population ($-$COHP) for all four structural models, where bonding and antibonding regions are indicated with a positive `+' and negative `$-$' sign, respectively, on the y-axis. A dashed vertical line indicates the Fermi level ($E_\text{F}$), defined as being at the middle of the bandgap.}
     \label{cohp}
     \refstepcounter{suppfigure}
 \end{figure*}

In Fig.~\ref{cohp}(a), we show the results of a crystal orbital Hamilton population (COHP) analysis to measure the bond strength, based on orbital interactions.\cite{Dronskowski-93-08, Deringer-11-06}
The $-$COHP bonding fingerprints are very similar across all four models, with a fully bonding contribution below $E_\text{F}$ (indicated by positive values in the plot), and a fully antibonding contribution (negative values) above $E_\text{F}$.

\begin{figure*}[h]
     \centering
     \includegraphics[]{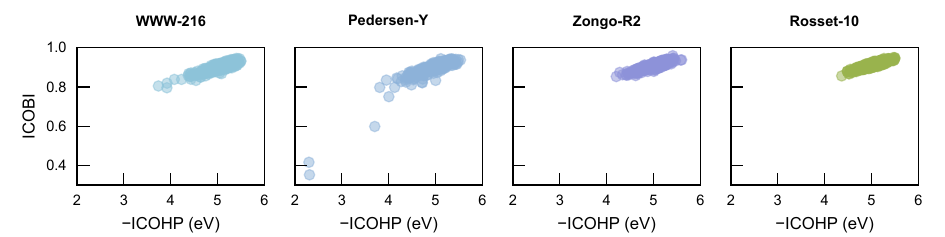}
     \caption{Scatter plots of the integrated crystal orbital bond index (ICOBI) against negative integrated crystal orbital Hamilton population ($-$ICOHP) for all four structural models.}
     \label{icohp}
     \refstepcounter{suppfigure}
 \end{figure*}

We plot ICOBI against ICOHP values for each bond in Fig.~\ref{icohp}, showing the relationship between the bond order (ICOBI) and the bond covalency ($-$ICOHP, in eV). The bond order increases with higher bond covalency in a linear fashion, excluding the bonds arising from coordination defects in the `Pedersen-Y' model.

\clearpage
\setstretch{1.0}

\section*{Supplemental references}